\documentclass[aps,prl,twocolumn,showpacs,groupedaddress]{revtex4}
\usepackage{graphics}
\begin{document}

\title{Transition between Orbital Orderings in YVO$_3$}

\author{G.R. Blake}
\author{T.T.M. Palstra}
\email[Corresponding author: ]{Palstra@chem.rug.nl}
\affiliation{Solid State Chemistry Laboratory, Materials Science Centre, University of Groningen,
 Nijenborgh 4, 9747 AG  Groningen, The Netherlands}
\author{Y. Ren}
\affiliation{Argonne National Laboratory, 9700 South Cass Avenue, Argonne, IL 60439}
\author{A.A. Nugroho}
\altaffiliation{On leave from Jurusan Fisika, Institut Teknologi Bandung, Jl. Ganesha 10, Bandung 40132, Indonesia.}
\author{A.A. Menovsky}
\affiliation{Van der Waals-Zeeman Institute, University of Amsterdam,
 Valckenierstraat 65, 1018 XE  Amsterdam, The Netherlands} 

\begin{abstract}
Evidence has been found for a change in the ordered occupation of the vanadium $d$-orbitals at the 77K phase transition in YVO$_3$, manifested by a change in the type of Jahn-Teller distortion. The orbital ordering above 77K is not destroyed at the magnetic ordering temperature of 116K, but is present as far as a second structural phase transition at 200K. The transition between orbital orderings is caused by an increase in octahedral tilting with decreasing temperature.
\end{abstract}

\pacs{61.10.Nz, 61.12.Ld, 61.50.Ks, 71.70.Ej}

\maketitle

The transition metal perovskite oxides $AB$O$_3$ form the basis for many interesting physical phenomena such as high $T_c$ superconductivity, ferroelectricity and colossal magnetoresistance. Despite these materials being the subject of intense study, in many cases their crystal structures have not been investigated in detail. For insulating materials containing a $B$-cation with orbital degeneracy such as Ti$^{3+}$, V$^{3+}$, Mn$^{3+}$ and Ni$^{3+}$, long-range ordering of the occupied $d$-orbitals is expected to take place below a transition temperature and will be accompanied by a Jahn-Teller (JT) distortion. Among the undoped perovskite oxides, JT distortions have been reported only for LaMnO$_3$ \cite{Rod98}, LaVO$_3$ \cite{Bor93}, GdTiO$_3$ \cite{MacL79} and YTiO$_3$ \cite{MacL79,Hes97}. However, orbital ordering can have profound and hitherto unrecognised effects on physical properties, such as the insulator-metal transition and charge order in doped La$_{1-x}M_x$MnO$_3$ \cite{vanAken01,End99}.

In this letter we report on the orbital ordering in YVO$_3$. The orbital ordering of $t_{2g}$ electron systems is reflected structurally to a lesser extent than that of $e_g$ systems but can still have a dramatic influence on the physical properties. We show that the orbital ordering in YVO$_3$ takes place at 200K which, contrary to previous assumptions, is far above the antiferromagnetic (AFM) ordering temperature $T_N=116$K. Moreover, there is a change in symmetry of the orbital ordering at $T_S=77$K. This is induced by an increased tilting of the octahedra with decreasing temperature, and it changes the easy axis of the V$^{3+}$ $d^2$ $S=1$ spin, resulting in a magnetic structure transition and a reversal of the net ferromagnetic moment of the canted AFM state \cite{Ren98}. 

We have previously described the unusual magnetic properties of YVO$_3$ \cite{Ren98,Ren00}. YVO$_3$ adopts a distorted perovskite structure with the space group $Pbnm$ \cite{sgroups} at room temperature. At $T_S$ there is a first order structural phase transition involving a sudden change in the unit cell volume, below which a JT ordered state is present \cite{Ren00}. A tetragonal distortion of the octahedra, where the long and short V-O bond distances alternate along the [110] and [1$\bar{1}$0] directions of the $ab$ plane, causes a splitting of the V $t_{2g}$ orbitals into a doublet of lower energy and a singlet of higher energy. The doublet contains the $xy$ orbital, which is always occupied, and either the $zx$ or $yz$ orbital. This ordered, alternating occupation of the V $d_{zx}$ and $d_{yz}$ orbitals between adjacent cations is shown schematically in Figs. 1 and 2. We believe that the change in magnetic structure \cite{Kaw94} is caused by a change from $C$-type orbital ordering (OO) below $T_S$ to $G$-type OO above $T_S$ (Fig. 2). This statement is supported by the Goodenough-Kanamori rules \cite{Good63}, band structure calculations \cite{Sawad98} and model Hartree-Fock calculations \cite{Miz99}. Experimentally, the only suggestion of $G$-type OO thus far has been provided by resonant X-ray scattering studies at the vanadium main $K$ edge \cite{Nog00}, although there is some debate about whether orbital ordering can be directly observed at this energy \cite{Elfimov99,Benfatto99}. In terms of the crystal structure, $G$-type OO is incompatible with $Pbnm$ symmetry since two crystallographically distinct JT-distorted $ab$ planes are required, with "out of phase" bonding arrangements; in $Pbnm$ all $ab$ planes are rendered equivalent by the mirror planes at $z=1/4$ and $3/4$, and the bonding arrangement is "in phase". We provide here the first crystallographic evidence for $G$-type OO above $T_S$.

Single crystals of YVO$_3$ were prepared as previously described \cite{Ren00}. Time-of-flight neutron powder diffraction data were collected on the instrument POLARIS at the ISIS facility using pulverised single crystals. The data were analysed by the Rietveld method as implemented in the GSAS program suite \cite{gsas}. Synchrotron X-ray diffraction was performed on beamline ID11 at ESRF, on a small single crystal fragment of approximate dimension 0.01mm at an X-ray energy of 42keV. Structural determination and refinement was carried out using the SHELXTL package \cite{shelx}.

\begin{figure}
\centering
\includegraphics{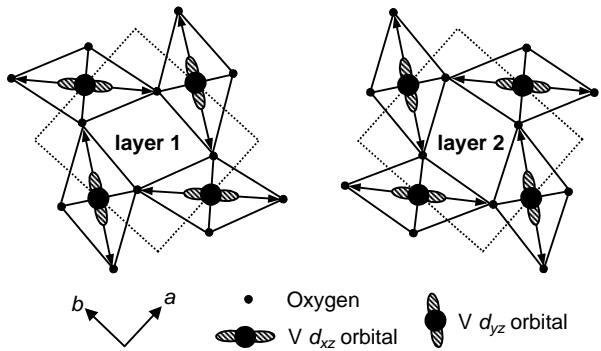}
\caption{Schematic picture of JT distortion in the $ab$ plane of YVO$_3$ below 200K; the unit cell is marked with a dotted line. Occupied V $d_{zx}$ and $d_{yz}$ orbitals are shown ($x$, $y$ and $z$ refer to nominal cubic perovskite axes); $d_{xy}$ orbitals are always occupied. There are two possible arrangements of alternating long (marked with arrows) and short V-O bonds, forming layers 1 and 2.} 
\end{figure}

\begin{figure}
\centering
\includegraphics{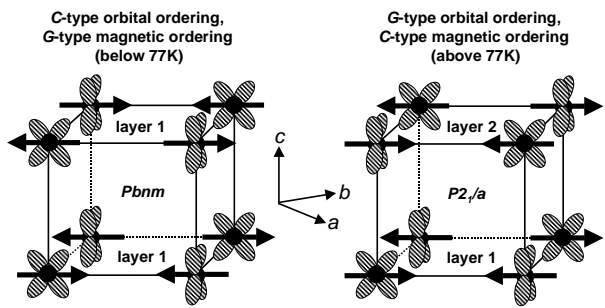}
\caption{Schematic representation of relationship between AFM ordering, orbital ordering and space group symmetry found in YVO$_3$. Occupied $d_{zx}$ and $d_{yz}$ orbitals are shown. Arrows represent AFM-ordered moments (arbitrary directions). The V-O bonding arrangement in successive layers (see Fig. 1) is either "in phase" ($Pbnm$) or "out of phase" ($P2_1/a$).}
\end{figure}

The starting model for Rietveld refinements of the neutron diffraction data was the 145K structure described by Nakotte \emph{et al.} \cite{Nak99}, a GdFeO$_3$-type distorted perovskite with space group $Pbnm$. Refinements at all temperatures proceeded smoothly. There is a clear phase transition between 65K and 80K; at 65K one long (2.043\AA) and two short (1.991\AA) V-O distances were obtained, the long and short bonds being arranged in an alternating pattern in the $ab$ plane, indicating the presence of a JT distortion corresponding to $C$-type OO. At $T\geq80$K three inequivalent V-O distances ranging from 1.985\AA\ to 2.027\AA\ were obtained, bonds within the $ab$ plane differing by only 0.01\AA. Refinements were also carried out in the monoclinic space group $P2_1/a$ \cite{sgroups}, using the LaVO$_3$ structure as a starting model, although no peaks were visibly split since $\beta$ remained close to 90$^{\circ}$. The refined structures at 65K, 240K and 295K were essentially the same in both space groups, with $\beta$ equal to 90$^{\circ}$ within a standard deviation. However, at intermediate temperatures there were significant structural differences between the refined $Pbnm$ and $P2_1/a$ models. In $P2_1/a$ the V-O bond lengths were split into one long (2.041\AA\ to 2.060\AA) and two short (1.982\AA\ to 2.001\AA) distances, the former being in the $ab$ plane. The bonding pattern was "out of phase" in successive planes, suggesting the presence of $G$-type OO. The $P2_1/a$ space group is compatible with both types of JT distortion, and at 65K the refined structure corresponded to $C$-type OO, with "in phase" bonding patterns in successive planes. However, there was little difference at any temperature between the quality of fits obtained using the two space groups; there was no concrete evidence to prove that the symmetry is ever lower than $Pbnm$.

The inconclusive structural refinement results from the powder diffraction study led us to continue our investigation using synchrotron X-ray single crystal diffraction. It soon became apparent that many weak reflections violating $Pbnm$ symmetry appeared on cooling through 200K. We note that both a change in slope of the magnetic susceptibility and an anomaly in the thermal expansion were previously observed at this temperature \cite{Ren00}. The intensity of one such forbidden reflection (401) is plotted as a function of temperature in Fig. 3. This provides definitive evidence for a phase transition with a lowering of the symmetry at approximately 200K. The forbidden reflections are very weak compared to those allowed in $Pbnm$, being of the order of 10$^{-4}$ times the intensity, and are therefore almost impossible to detect using a conventional laboratory X-ray source.

The data sets collected at 295K, 225K and 50K were consistent with the space group $Pbnm$ and did not contain any extra reflections; the refinements gave structures very close to those obtained from the neutron diffraction study at similar temperatures. A strong JT distortion corresponding to $C$-type OO was present in the 50K structure. However, the data sets collected at between 80K and 175K contained many extra, weak reflections; the reflection conditions were consistent with the space group $P2_1/a$. The introduction of a pseudo-merohedral twinning model involving domains rotated with respect to each other by 180$^{\circ}$ around the $a$ or $c$ axis (both rotations are equivalent in $P2_1/a$) improved the refinements considerably. The $ab$ planes in all of the refined structures between 80K and 175K contained the expected alternating pattern of long and short V-O bonds. The bonding patterns in successive planes were "out of phase", characteristic of $G$-type OO.

The structures obtained from the synchrotron and neutron data refinements were very similar; the yttrium positions were determined rather more precisely from the synchrotron data but the lattice parameters, oxygen coordinates and thermal factors were better determined from the neutron data. In this case the two techniques are complementary; the synchrotron single crystal data are essential in order to detect the subtle lowering of symmetry, while the neutron powder data yield more accurate values for most structural parameters if the correct space group is known. The lattice parameters determined by neutron diffraction are given in Table I, atomic coordinates at three representative temperatures of 65K, 140K and 295K are listed in Tables II, III and IV, and selected bond lengths and angles are listed in Table V.

\begin{figure}
\centering
\includegraphics{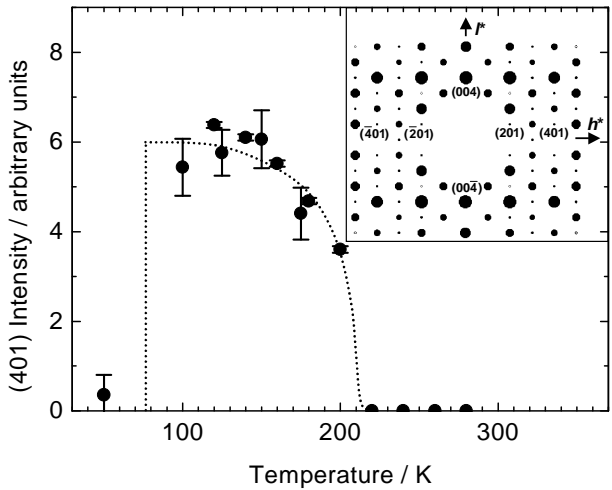}
\caption{Temperature dependence of the "forbidden" (401) reflection measured by synchrotron X-ray diffraction. Intensities with large error bars were obtained from the automatic integration of an entire data file; intensities with small error bars were integrated individually. Inset shows a section through reciprocal space obtained at 175K where the (401) and other "forbidden" ($h$0$l$), $h+l\neq2n$ reflections are apparent; intensities of reflections are represented by sizes of circles.} 
\end{figure}

The V-O bond lengths obtained using the neutron data are plotted as a function of temperature in Fig. 4. At temperatures above the 200K phase transition there are three unequal V-O bond distances, while below 200K there is a clear JT distortion, the bonds being split into one long ($\sim2.05$\AA) and two short ($\sim1.99$\AA) distances in both orbitally ordered temperature regimes, the longer distance always being in the $ab$ plane. The difference between the longest and shortest bond lengths ($\sim0.06$\AA) is small compared to that in an $e_g$ orbitally ordered system such as LaMnO$_3$ ($\sim0.27$\AA) \cite{Rod98} but similar to that present in LaVO$_3$ ($\sim0.06$\AA) \cite{Bor93}. Differences in bond lengths of similar magnitude occur in the room temperature structures of YTiO$_3$ and GdTiO$_3$, which both contain Ti$^{3+}$ cations where the $t_{2g}$ orbitals are occupied by a single electron \cite{MacL79,Hes97}. In YTiO$_3$, GdTiO$_3$ and LaMnO$_3$ the onset of JT distortion also occurs at temperatures far above the onset of magnetic ordering; the magnetic and orbital ordering temperatures in perovskites appear to be completely independent. Among the orthovanadates, the AFM ordering temperature steadily decreases from 144K for LaVO$_3$ \cite{Ngu95a} to 101K for LuVO$_3$ \cite{Zub76a} as the lanthanide cation decreases in size from left to right across the series. This is due to an increase in the degree of octahedral tilting, which leads to the vanadium 3$d$ electrons becoming more localised and the V-O-V superexchange interactions becoming weaker. The general trend of the orbital ordering temperature in $AB$O$_3$ perovskites across the lanthanide series, as manifested by the onset of JT distortion, has been little investigated; among the orthovanadates the JT ordering temperature is only known for LaVO$_3$ (between 140K and 145K \cite{Bor93}) and now YVO$_3$ (200K). This is perhaps due to the difficulty in recognising $G$-type OO in these two materials (and which is probably present in all the orthovanadates from LaVO$_3$ to DyVO$_3$); it is incompatible with $Pbnm$ symmetry and can only be detected if a lowering of the symmetry to monoclinic is observed. In the case of YVO$_3$ the unit cell remains metrically orthorhombic, and the reflections violating $Pbnm$ in diffraction experiments are so weak that single crystals and high intensity synchrotron radiation are required in order to detect the phase transition involved. An additional complication is the pseudo-merohedral twinning inherent in perovskite single crystals that undergo an orthorhombic to monoclinic phase transition; it does not lead to the appearance of additional reflections and can be difficult to identify.

\begin{figure}
\centering
\includegraphics{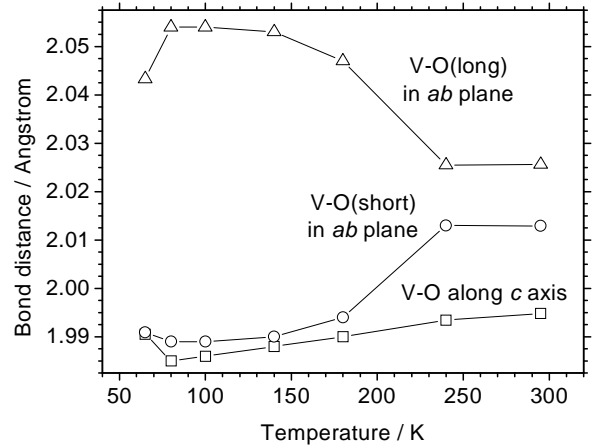}
\caption{Temperature dependence of V-O bond distances determined using neutron powder diffraction. For the $P2_1/a$ structures, the "equivalent" V-O distances in successive layers 1 and 2 (see Fig. 1) have been averaged. Typical error bars are 0.001\AA.} 
\end{figure}

The yttrium position determined from the synchrotron refinements is plotted as a function of temperature in Fig. 5a. The $z$ coordinate has been neglected since it is fixed by symmetry for the $Pbnm$ structures and deviates very little from 1/4 or 3/4 for the $P2_1/a$ structures. The yttrium cation does not remain in the same position relative to the fixed vanadium cations but moves steadily in the direction shown in Fig. 5a as the temperature is lowered to 80K. Below $T_S$ the shift appears to change direction, becoming close to [100]. The magnitude of the shift relative to the 295K position (Fig. 5a, inset) is of the order of 0.008\AA\ between 295K and 80K, with a disproportionately large increase on cooling through $T_S$. This movement of the $A$-site cation is probably caused by an increase in the degree of octahedral tilting \cite{Miz99}. The V-O-V bond angle between adjacent octahedra along the $c$ axis, as obtained from the neutron diffraction study, is plotted as a function of temperature in Fig. 5b; it decreases over the temperature range studied and mirrors the trend of the yttrium site shift. For perovskites in general, the displacement of oxygen atoms from their ideal positions due to octahedral tilting will tend to push the $A$-site cation in a given direction, as explained in detail by Mizokawa \emph{et al.} \cite{Miz99}. The driving force for this shift is the maximisation of $A$-O covalency (hybridisation between the $A$-site cation $d$-orbitals and the oxygen ion 2$p$ orbitals); it is energetically favourable for the $A$-site cation to remain close to four nearest neighbour oxygen ions, while little energy is lost in moving further away from the oxygen ions situated at a greater distance. This is the case in YVO$_3$, where there are four Y-O distances of between 2.24\AA\ and 2.30\AA, the next shortest distance being $\sim$2.49\AA. The $C$-type OO is stabilised by a high degree of octahedral tilting because two of the four closest oxygen ions, one in each of two successive $ab$ planes, are moved in approximately the same direction by the corresponding JT distortion, allowing the $A$-site cation to be pushed in a given direction while retaining close contact with all four oxygen ions. 
However, when $G$-type OO is present, the oxygen ion shifts in successive $ab$ planes are "out of phase" with respect to each other and the $A$-site cation is pushed in two conflicting directions at once, decreasing the energy gain due to the $A$-O hybridisation energy. It follows that $C$-type OO is predicted to be most stable for perovskites with a large degree of octahedral tilting, while $G$-type OO is more stable for lesser degrees of tilting, where the energy gain from the orbital ordering is greater than that from the $A$-O hybridisation \cite{Miz99}. The type of AFM ordering is determined by the type of orbital ordering, if present, but the onset temperature of each is independent. In going from left to right across the lanthanide orthovanadate series, it has been reported that a change from ground-state $C$-type to $G$-type AFM ordering takes place between DyVO$_3$ and HoVO$_3$ as the octahedral tilting increases \cite{Zub76a}; this agrees with the above argument. A simultaneous transition from ground-state $G$-type to $C$-type OO is almost certainly responsible for the crossover in AFM structures, but has not been investigated. The Y$^{3+}$ cation is very similar in size to Ho$^{3+}$, and YVO$_3$ appears to be a unique material with octahedral tilting of a critical magnitude, at which both possible orbital ordering configurations are of similar energies. The slight increase in the degree of octahedral tilting with falling temperature causes a transition from $G$-type to $C$-type OO on cooling through $T_S$. The rearrangement of the occupied orbitals in turn forces a change in the type of AFM configuration and in the direction of the easy axis, leading to a reversal of the net ferromagnetic moment. 

\begin{figure}
\centering
\includegraphics{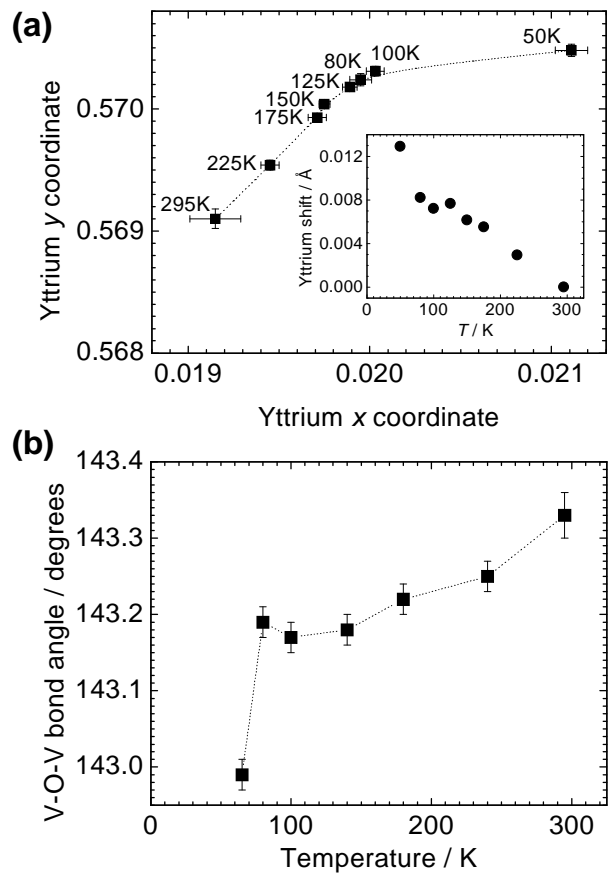}
\caption{(a.) Variation of yttrium $x$ and $y$ coordinates with temperature; inset shows magnitude of Y-shift (zero point taken as 295K position). (b.) Temperature dependence of V-O-V angle along $c$ axis.} 
\end{figure}

The above discussion is most likely applicable to JT-distorted perovskites in general. The degree of octahedral tilting is the decisive factor in determining the most stable orbital ordering configuration, which in turn determines the magnetic structure at temperatures below $T_N$. It is likely that other JT-active $AB$O$_3$ perovskites with a critical $A$:$B$ size ratio will display magnetisation reversals like that at $T_S$ in YVO$_3$. It is therefore worth studying related perovskites more closely to gain further insight into the interplay between octahedral tilting, orbital ordering and electronic properties.

\begin{acknowledgments}
We thank R.I. Smith and G. Vaughan for experimental assistance at ISIS and ESRF respectively. We are grateful to G.A. Sawatzky, D.I. Khomskii, A. Meetsma and B.B. van Aken for useful discussions. This work was supported by the EU via the TMR Oxide Spin Electronics Network (OXSEN) and by the Netherlands Organisation for Scientific Research (NWO).
\end{acknowledgments}

\begin{widetext}

\begin{table}
\centering \caption{Lattice parameters for YVO$_3$}
\begin{ruledtabular}
\begin{tabular}{c c c c c c c}
$T$/K&Space group&$a$ / \AA&$b$ / \AA&$c$ / \AA&$\beta$ / $^{\circ}$&Volume / \AA$^3$\\ \hline \\
295&$Pbnm$&5.27839(3)&5.60608(3)&7.57421(4)&&224.129(1)\\
240&$Pbnm$&5.27574(3)&5.60679(3)&7.56714(4)&&223.836(1)\\
180&$P2_1/a$&5.61126(3)&5.27474(4)&7.55316(4)&89.978(3)&223.558(1)\\
140&$P2_1/a$&5.61602(3)&5.27393(3)&7.54235(4)&89.973(3)&223.393(1)\\
100&$P2_1/a$&5.61940(3)&5.27272(3)&7.53499(4)&89.979(3)&223.258(1)\\
80&$P2_1/a$&5.62058(3)&5.27243(3)&7.53254(4)&89.977(3)&223.220(1)\\
65&$Pbnm$&5.28164(3)&5.58868(3)&7.55030(4)&&222.865(1)\\
\end{tabular}
\end{ruledtabular}
\end{table}

\begin{table}
\centering \caption{Atomic parameters for YVO$_3$ at 65K}
\begin{ruledtabular}
\begin{tabular}{c c c c c c c c c c c}
Atom&Position&$x$&$y$&$z$&$U_{11}$ or $U_{iso}$&$U_{22}$&$U_{33}$&$U_{12}$&$U_{13}$&$U_{23}$\\ \hline \\
Y&4$c$&0.97859(6)&0.42944(5)&3/4&0.0024(1)&0.0020(1)&0.0020(1)&0.0001(1)&0&0\\
V&4$a$&0&0&0&0.003&&&&&\\
O1&4$c$&0.88757(8)&0.96140(7)&1/4&0.0030(1)&0.0041(1)&0.0024(1)&0.0002(1)&0&0\\
O2&8$d$&0.18691(5)&0.29958(5)&0.05710(4)&0.0031(1)&0.0039(1)&0.0036(1)&-0.0005(1)&0.0007(1)&-0.0009(1)\\
\end{tabular}
\end{ruledtabular}
($wRp=0.0142$, $\chi^2=1.836$)
\end{table}

\begin{table}
\centering \caption{Atomic parameters for YVO$_3$ at 140K}
\begin{ruledtabular}
\begin{tabular}{c c c c c c c c c c c}
Atom&Position&$x$&$y$&$z$&$U_{11}$ or $U_{iso}$&$U_{22}$&$U_{33}$&$U_{12}$&$U_{13}$&$U_{23}$\\ \hline \\
Y&4$e$&0.07042(6)&0.97985(6)&0.2493(2)&0.0031(1)&0.0034(1)&0.0029(1)&-0.0001(1)&-0.0018(2)&-0.0010(3)\\
V1&2$c$&1/2&0&0&0.004&&&&&\\
V2&2$b$&0&1/2&1/2&0.004&&&&&\\
O1&4$e$&0.46037(7)&0.11126(7)&0.2508(4)&0.0048(1)&0.0038(1)&0.0028(1)&-0.0007(1)&-0.0008(5)&0.0006(4)\\
O2&4$e$&0.3087(2)&0.6946(2)&0.0557(3)&0.0047(3)&0.0040(3)&0.0036(4)&0.0005(2)&-0.0008(3)&0.0016(3)\\
O3&4$e$&0.6999(2)&0.3140(2)&0.5576(3)&0.0032(3)&0.0030(3)&0.0047(4)&-0.0015(2)&0.0011(3)&0.0000(3)\\
\end{tabular}
\end{ruledtabular}
($wRp=0.0128$, $\chi^2=1.549$)
\end{table}

\begin{table}
\centering \caption{Atomic parameters for YVO$_3$ at 295K}
\begin{ruledtabular}
\begin{tabular}{c c c c c c c c c c c}
Atom&Position&$x$&$y$&$z$&$U_{11}$ or $U_{iso}$&$U_{22}$&$U_{33}$&$U_{12}$&$U_{13}$&$U_{23}$\\ \hline \\
Y&4$c$&0.98071(7)&0.43061(6)&3/4&0.0058(1)&0.0050(1)&0.0049(1)&0.0005(1)&0&0\\
V&4$a$&0&0&0&0.005&&&&&\\
O1&4$c$&0.88885(9)&0.96031(9)&1/4&0.0053(2)&0.0073(2)&0.0040(1)&0.0013(1)&0&0\\
O2&8$d$&0.19084(6)&0.30408(6)&0.05640(4)&0.0054(1)&0.0064(1)&0.0063(1)&-0.0009(1)&0.0015(1)&-0.0019(1)\\
\end{tabular}
\end{ruledtabular}
($wRp=0.0128$, $\chi^2=1.488$)
\end{table}

\begin{table}
\centering \caption{Selected bond distances (\AA) and angles (degrees) for YVO$_3$}
\begin{ruledtabular}
\begin{tabular}{c c c c c c c c}
&295K ($Pbnm$)&240K ($Pbnm$)&180K ($P2_1/a$)&140K ($P2_1/a$)&100K ($P2_1/a$)&80K ($P2_1/a$)&65K ($Pbnm$)\\ \hline \\
Y-O1&2.2395(6)&2.2401(5)&2.2408(5)&2.2433(5)&2.2433(5)&2.2445(5)&2.2446(5)\\
Y-O1&2.2971(6)&2.2960(5)&2.2967(5)&2.2971(5)&2.2996(5)&2.3001(5)&2.2958(5)\\
Y-O2&2.2765(4)$\times2$&2.2749(4)$\times2$&2.267(2)&2.267(2)&2.266(2)&2.266(2)&2.2758(3)$\times2$\\
Y-O2&2.4945(4)$\times2$&2.4910(4)$\times2$&2.487(2)&2.487(2)&2.485(3)&2.484(3)&2.4764(4)$\times2$\\
Y-O2&2.6681(4)$\times2$&2.6668(4)$\times2$&2.656(3)&2.652(3)&2.655(3)&2.657(3)&2.6671(3)$\times2$\\
Y-O3&&&2.281(2)&2.278(2)&2.278(2)&2.278(2)&\\
Y-O3&&&2.492(2)&2.488(2)&2.487(2)&2.485(3)&\\
Y-O3&&&2.669(3)&2.668(3)&2.662(3)&2.661(3)&\\

V1-O1&1.9948(2)$\times2$&1.9934(1)$\times2$&1.997(3)$\times2$&1.993(3)$\times2$&1.986(3)$\times2$&1.987(3)$\times2$&1.9905(1)$\times2$\\
V1-O2&2.0129(3)$\times2$&2.0130(3)$\times2$&1.987(1)$\times2$&1.981(1)$\times2$&1.981(1)$\times2$&1.981(1)$\times2$&1.9909(3)$\times2$\\
V1-O2&2.0256(4)$\times2$&2.0255(3)$\times2$&2.052(1)$\times2$&2.058(1)$\times2$&2.060(1)$\times2$&2.060(1)$\times2$&2.0433(3)$\times2$\\
V2-O1&&&1.983(3)$\times2$&1.982(3)$\times2$&1.985(3)$\times2$&1.982(3)$\times2$&\\
V2-O3&&&2.001(1)$\times2$&1.998(1)$\times2$&1.997(1)$\times2$&1.997(1)$\times2$&\\
V2-O3&&&2.041(1)$\times2$&2.047(1)$\times2$&2.048(1)$\times2$&2.047(1)$\times2$&\\
\\
V1-O-V1(V2)&143.33(3)&143.25(2)&143.22(2)&143.18(2)&143.17(2)&143.19(2)&142.99(2)\\
V1-O2-V1&144.85(2)&144.79(2)&144.87(9)&144.98(9)&144.88(10)&144.87(11)&144.75(2)\\
V2-O3-V2&&&144.60(10)&144.46(10)&144.57(11)&144.65(11)&\\
\\
O1-V1-O2&87.71(2)&87.70(2)&87.49(7)&87.31(7)&87.36(8)&87.41(8)&88.07(1)\\
O1-V1-O2&88.63(2)&88.64(1)&88.66(7)&88.74(7)&88.70(8)&88.65(8)&88.97(1)\\
O1-V1-O2&91.37(2)&91.36(2)&91.34(7)&91.26(7)&91.30(8)&91.35(8)&91.03(1)\\
O1-V1-O2&92.29(2)&92.30(2)&92.51(7)&92.69(7)&92.64(8)&92.59(8)&91.93(1)\\
O2-V1-O2&89.352(6)&89.325(5)&89.55(3)&89.53(3)&89.54(4)&89.53(4)&89.197(5)\\
O2-V1-O2&90.648(6)&90.675(5)&90.45(3)&90.47(3)&90.46(4)&90.47(4)&90.803(5)\\
O1-V2-O3&&&87.94(7)&88.03(7)&88.02(8)&87.97(8)&\\
O1-V2-O3&&&88.73(7)&88.65(7)&88.74(8)&88.75(8)&\\
O1-V2-O3&&&91.27(7)&91.35(7)&91.26(8)&91.25(8)&\\
O1-V2-O3&&&92.06(7)&91.97(7)&91.98(8)&92.03(8)&\\
O3-V2-O3&&&88.98(3)&88.89(3)&88.80(4)&88.78(4)&\\
O3-V2-O3&&&91.02(3)&91.11(3)&91.20(4)&91.22(4)&\\
\end{tabular}
\end{ruledtabular}
\end{table}
\end{widetext}

\begin{thebibliography}{21}
\bibitem[{sgr()}]{sgroups}
\bibinfo{title}{The unit cell of $a\approx5.61$, $b\approx7.57$,
  $c\approx5.28$\AA\ in the standard setting $Pnma$ is transformed to
  $a\approx5.28$, $b\approx5.61$, $c\approx7.57$\AA\ in $Pbnm$; most of the
  listed citations use $Pbnm$, which we retain for consistency. We use $P2_1/a$
  for the monoclinic phase instead of the standard setting $P2_1/c$ in order to
  retain the doubled perovskite $c$ axis.}
\end{thebibliography}
\end{document}